\documentclass[10 pt]{article}
\usepackage[cp1250]{inputenc}
\usepackage{color}
\usepackage{amsfonts, amssymb, amsthm, amsmath}
\usepackage{epsfig}
\usepackage{latexsym}
\begin{document}
\title{Heat, temperature and relativity}
\author{Maciej Przanowski\\
\small{Institute of Physics, Technical University of \L\'{o}d\'{z}, W\'{o}lcza\'{n}ska 219, 90-924  \L\'{o}d\'{z}}}

\maketitle
\begin{center}
{\it Dedicated to Professor Leszek Wojtczak}\\[2 ex]
\mbox{ }
\end{center}
The present work is motivated by still ongoing controversies on such fundamental concepts of relativistic 
thermodynamics as heat, work, temperature or probability (see e.g. \cite{1} and references therein, and \cite{2}).
We are going to show how a distinguished paper by Cubero et al \cite{3} sheds new light on these controversies 
and enables us to solve them to some extent. In \cite{3} the authors present their results on one-diemensional
relativistic particle dynamics simulations. They have shown that numerical results of the simulations are in 
excellent agreement  with the {\it J\H{u}ttner distribution} \cite{4} if the termodynamic system (ideal gas) 
is at rest with respect to the inertial laboratory frame or with some counterpart of J\H{u}ttner distribution
if this gas moves with a constant velocity. Assuming that the same is true in three dimensions one finds that 
the relativistic generalization of the Maxwell distribution for ideal gas is given by the {\it J\H{u}ttner 
distribution} \cite{4}
\begin{equation}
\label{Juttner1}
dw_{\vec{p}}^{(J)}=\frac{1}{Z}\exp\{-\beta\, c\sqrt{\vec{p}^{2}+m^{2}c^{2}}\}dp_{x}dp_{y}dp_{z}
\end{equation}
if the velocity of the gas $\vec{V}=0$, and for any $\vec{V}$ by
\begin{equation}
\label{Juttner2}
dw_{\vec{p}}=\frac{1}{Z\,\gamma(\vec{V})}\exp\{-\beta\, c\,u_{j}p^{j}\}dp_{x}dp_{y}dp_{z}
\end{equation}
where $Z$ is the normalization constant,  $\vec{p}=(p_{x},p_{y},p_{z})$ is momentum 
of the particle; 
$p^{j}$, $j=1,2,3,4$, stands for the four-vector of momentum i.e., 
$p^{j}=(\vec{p},\frac{{\mathcal E}}{c})\, ,\hspace{1 ex}{\mathcal E}=c\sqrt{\vec{p}^{2}+m^{2}c^{2}};$\hspace{1 ex}
$u^{j}$ is the four-velocity of thermodynamic system i.e., $u^{j}=(\gamma(\vec{V})\frac{\vec{V}}{c},\gamma(\vec{V}))$,
\hspace{1 ex} $\gamma(\vec{V})=\frac{1}{\sqrt{1-\frac{\vec{V}^{2}}{c^{2}}}}$; finally $\beta=\dfrac{1}{k\,T_{0}}$, 
where $k$ is the Boltzmann constant and $T_{0}$ is the gas temperature in the rest frame of the gas.

From (\ref{Juttner2}) we quickly infer that the relativistic Gibbs distribution reads
\begin{equation}
\label{Gibbs_relativ}
dw=\frac{V^{N}}{(2\pi\hbar)^{3N}N!{\mathcal Z}}\exp\{-\beta\, c\,u_{j}P^{j}\}d^{3N}p
\end{equation}
where $V$ is the volume of the system, $N$ denotes the number of particles, $P^{j}=(\vec{P},\frac{E}{c})$
is the total four-momentum and ${\mathcal Z}$ stands for the partition function
\begin{equation}
\label{relativ_Z}
{\mathcal Z}=\frac{V^{N}}{(2\pi\hbar)^{3N}N!}\int\exp\{-\beta\, c\,u_{j}P^{j}\}d^{3N}p
\end{equation}
Then the entropy $S$ reads
\begin{equation}
\label{entropy1}
S=-k\langle \ln(\frac{1}{{\mathcal Z}}\exp\{-\beta\, c\,u_{j}P^{j}\})\rangle = 
k(\ln {\mathcal Z}+\beta c u_{j}\langle P^{j}\rangle)
\end{equation}
and it can be shown by direct calculations (see for example \cite{5}) that  the partition function ${\mathcal Z}$, 
the entropy $S$ and the pressure $P$ of the gas are relativistic invariants. Moreover, straightforward calculations 
lead to the relations \cite{1},\cite{5,6,7,8,9}
\begin{equation}
\label{avarage_P}
\langle E\rangle+PV=\gamma(\vec{V})(\langle E_{0}\rangle+P_{0}V_{0}),\hspace{3 ex}
\langle \vec{P}\rangle=\gamma(\vec{V})(\langle E_{0}\rangle+P_{0}V_{0})\frac{\vec{V}}{c^{2}}
\end{equation}
which say that $(\langle\vec{P}\rangle,\dfrac{\langle E\rangle+PV}{c})$ constitutes a four-vector. 
Consequently, $(\langle \vec{P}\rangle,\dfrac{\langle E\rangle}{c})$ is not a four-vector, but according to 
(\ref{entropy1}) $u_{j}\langle P^{j}\rangle$ is a Lorentz invariant. In the paper the subindex "$_{0}$" stands
for the physical quantities in the rest frame of the system.

From (\ref{avarage_P}) one gets immedietly the first law of thermodynamics in special relativity in the form
\begin{equation}
\label{1st_law}
d\langle P^{j}\rangle=\frac{1}{c}T_{0}u^{j}\,dS\, -\, \frac{1}{c}\delta^{j}_{\, 4}\, d(PV)\,+\,
\frac{1}{c}\gamma(\vec{V})Vu^{j}\,dP\,+\,\frac{\langle E_{0}\rangle+P_{0}V_{0}}{c}du^{j}
\end{equation}
Hence, we conclude from (\ref{1st_law}) that the four-vectors of temperature $T^{j}$ and heat $\delta Q^{j}$
are defined by
\begin{equation}
\label{definition_T}
T^{j}:=\frac{1}{c}T_{0}u^{j}\, ,\hspace{3 ex}\delta Q^{j}:=T^{j}\,dS
\end{equation}
and the four-object of relativistic work $\delta L^{j}$ (which is not a four-vector!) reads
\begin{equation}
\label{work1}
\delta L^{j}=-\, \frac{1}{c}\delta^{j}_{\, 4}\, d(PV)\,+\,
\frac{1}{c}\gamma(\vec{V})Vu^{j}\,dP\,+\,\frac{\langle E_{0}\rangle+P_{0}V_{0}}{c}du^{j}
\end{equation}
Consequently, the temperature $T=cT_{4}$ and the heat $\delta Q=c\delta Q_{4}$ transform as follows
\begin{equation}
\label{10}
T=\gamma(\vec{V})T_{0}\, ,\hspace{3 ex}\delta Q=\gamma(\vec{V})\delta Q_{0}
\end{equation}
as has been proved by H. Ott \cite{10}, H. Arzeli\'{e}s \cite{11} and C. M\o ller \cite{8,9} (see also
\cite{1}, \cite{12}, \cite{13}). Then the work $\delta L=c\delta L_{4}$ takes the form
\begin{equation}
\label{11}
\delta L=-PdV+(\gamma(\vec{V}))^{2}\frac{\vec{V}^{2}}{c^{2}}VdP + (\langle E_{0}\rangle+P_{0}V_{0})d\gamma(\vec{V})
\end{equation}
(compare \cite{1}, Eq. (44)).

Now we find the first law of thermodynamics directly from the Gibbs distribution (\ref{Gibbs_relativ}). 
Standard calculations of statistical thermodynamics (see for example \cite{14}) give
\begin{equation}
\label{1st_law2}
u_{j}d\langle P^{j}\rangle=\frac{1}{kc\beta}dS\, +\, u_{j}\langle dP^{j}\rangle
\end{equation}
From (\ref{1st_law2}) one infers that $d\langle P^{j}\rangle-\langle dP^{j}\rangle$ is a four-vector, and 
\begin{equation}
\label{1st_law3}
d\langle P^{j}\rangle=\frac{1}{kc\beta}u^{j}dS\, +\, \langle dP^{j}\rangle\, +\, \delta Q_{\perp}^{j}
\end{equation}
where $\delta Q_{\perp}^{j}$ is a four-vector of heat orthogonal to $u^{j}$ i.e. 
\begin{equation}
u_{j}\delta Q_{\perp}^{j}=0.
\end{equation}
Assuming that  in the rest frame of gas, for any reversible process one has 
(see (\ref{work1}) for $\vec{V}=0$ and Ref. \cite{8})
\begin{equation}
\delta Q_{0\perp}^{4}=0\, ,\hspace{3 ex}\delta Q_{0\perp}^{\mu}=-\langle dP_{0}^{\mu}\rangle\,=0\, , \hspace{2 ex}
\mu=1,2,3,
\end{equation}
we rewrite (\ref{1st_law3}) in the following form
\begin{equation}
\label{1st_law4}
d\langle P^{j}\rangle=\frac{1}{kc\beta}u^{j}dS\, +\, \langle dP^{j}\rangle\,
\end{equation}
Equation (\ref{1st_law4})
is the first law of thermodynamics derived from the relativistic Gibbs distribution (\ref{Gibbs_relativ}). Therefore
\begin{equation}
\label{heat}
\delta Q^{j}=\frac{1}{kc\beta}u^{j}dS\hspace{3 ex}\Longrightarrow\hspace{3 ex}T^{j}=\frac{1}{kc\beta}u^{j}, \hspace{2 ex}
T=\frac{1}{k\beta}\gamma(\vec{V})
\end{equation}
according to (\ref{definition_T}) and (\ref{10}), and as usually
\begin{equation}
\delta L^{j}=\langle dP^{j}\rangle .
\end{equation}
Relativistic temperature $T$ given by (\ref{10}) can be measured with the use of relativistic Carnot cycle
\cite{9},\cite{1}. To see this we consider a thermodynamic engine being a slightly simplified version of the 
one analysed by M\o ller \cite{9}. The engine realizes the relativistic Carnot cycle and it operates 
between two reservoirs $R_{0}$ and $R$. The reservoir $R$ moves with a constant velocity $\vec{V}$ with 
respect to $R_{0}$. The temperature of both $R_{0}$ and $R$ in their rest frames is $T_{0}$ and the 
temperature of $R$ with respect to the rest frame of $R_{0}$ is, by (\ref{10}), $T=\gamma(\vec{V})T_{0}$.
The engine works as follows
\begin{enumerate}
\item[(I)] The amount of heat $Q_{0}$ is absorbed isothermically from $R_{0}$ at the temperature $T_{0}$.
\item[(II)] The system is accelerated adiabatically to the velocity $\vec{V}$.
\item[(III)] The amount of heat $Q=\gamma(\vec{V})Q_{0}$ (with respect to the rest frame of $R_{0}$!) is released 
isothermically from the system to $R$ at the temperature $T=\gamma(\vec{V})T_{0}$ (with respect to the rest 
frame of $R_{0}$!).
\item[(IV)] Finally, the system is decelerated adiabatically so that it returns to the initial state.
\end{enumerate}
From the second law of thermodynamics it follows that 
\begin{equation}
\label{13}
\frac{Q}{Q_{0}}=\frac{T}{T_{0}}=\gamma(\vec{V})
\end{equation}
and this {\it enables us to check experimentally the validity of the transformation rules} (\ref{10}). Observe
also that the efficiency of our Carnot cycle reads
\begin{equation}
\label{efficiency}
\eta=\dfrac{T_{0}-T}{T_{0}}=1-\gamma(\vec{V})\, <0.
\end{equation}
If the transformation rules were 
$T=(\gamma(\vec{V}))^{-1}T_{0}$,  $\delta Q=(\gamma(\vec{V}))^{-1}\delta Q_{0}$,\linebreak 
as was assumed by M. Planck, K.V. Mosengeil, M.V. Laue, W. Pauli or A. Einstein (who finally changed his opinion 
in 1952/53) (see \cite{1}), then the efficiency would be 
\begin{equation}
\label{efficiency2}
\eta=1-\frac{1}{\gamma(\vec{V})}\, >0.
\end{equation}

In contrary to phenomenological thermodynamics in relativistic statistical thermodynamics the transformation 
rule for temperature depends on the definition of "statistical thermometer" and there exists no natural, uniquely 
defined rule of transformation. For example if you assume that by comparing (\ref{Juttner2}) with (\ref{Juttner1}) 
the temperature should be defined as 
\begin{equation}
\label{efficiency2}
T:=\dfrac{1}{k\beta u^{4}}=\frac{1}{\gamma(\vec{V})}T_{0}
\end{equation}
you get the transformation rule of M. Planck, K.V. Mosengeil and others. Moreover, as can be easily shawn 
by using the Bose-Einstein counterpart of (\ref{Juttner2}) for black body radiation the temperature of a moving black
body is not well defined \cite{15},\cite{16}.

\noindent [Remark: After preparing this paper for publication I found Ref. \cite{17} where relativistic thermodynamics from the 
point of view of J\H{u}ttner distribution has been also considered.]


\thebibliography{99}
\bibitem{1} M.Requardt, Thermodynamics meets Special Relativity - or what is real in Physics, 
arXiv: 0801.2639v1 [gr-qc]
\bibitem{2} K.A. Johns and P.T. Landsberg, {\it J. Phys. A: Gen. Phys.} {\bf 3}, 113 (1971)
\bibitem{3} D. Cubero, J. Casado-Pascual, J. Dunkel, P. Talkner and P. H\H{a}nggi, 
{\it Phys. Rev. Lett.} {\bf 99}, 170601 (2007)
\bibitem{4} F. J\H{u}ttner, {\it Ann. Phys.} (Leipzig) {\bf 34}, 856 (1911) 
\bibitem{5} R.K. Pathria, {\it Proc. Nat. Inst. Sci. India} {\bf 23 A}, No 3, 168 (1957).
\bibitem{6} R.K. Pathria, {\it Proc. Nat. Inst. Sci. India} {\bf 23 A}, No 5, 331 (1955).
\bibitem{7} A. Staruszkiewicz, {\it Acta Phys. Polon.} {\bf 29}, 249 (1966)
\bibitem{8} C. M\o ller, Relativistic Thermodynamics (A strange incident in the History of Physics)
{\it Det. Kong. Danske Videnskab. Selskab. Mat.-fys. Medd.}, {\bf 36} (Kobenhavn 1967).
\bibitem{9} C. M\o ller, Thermodynamics in the Special and the General Theory of Relativity,
in {\it Bernardini Festschrift}, ed. G. Poppi (Academic Press, 1968) pp. 202-221.
\bibitem{10} H. Ott, {\it Zeitschr. d. Phus.} {\bf 175}, 70 (1963).
\bibitem{11} H. Arzeli\'{e}s, {\it Nuovo Cim.} {\bf 35}, 792 (1965).
\bibitem{12} N.G. van Kampen, {\it Phys. Rev.} {\bf 173}, 295 (1968).
\bibitem{13} D. Ter Haar and H. Wergeland {\it Phys. Rep.} {\bf 1}, 31 (1971)
\bibitem{14} L.D. Landau and E.M. Lifschitz, {\it Statistical Physics}, (Pergamon Press, Oxford 1969).
\bibitem{15} P.T. Landsberg and G.E.A. Matsas, {\it Phys. Lett.} A, {\bf 223}, 401 (1996)
\bibitem{16} P.T. Landsberg and G.E.A. Matsas, {\it Physica} A, {\bf 340}, 92 (2004)
\bibitem{17} J. Dunkel, P. H\H{a}nggi and S. Hilbert, Nonlocal oservables and lightcone-averaging 
in relativistic thermodynamics, arXiv: 0902.4651v2 [cond-mat.stat-mech]
\end{document}